\documentclass[aps,prl,reprint,showpacs]{revtex4-1} 
\usepackage[utf8]{inputenc}
\usepackage{graphicx}
\usepackage{amsmath}
\usepackage{amssymb}
\usepackage{verbatim}
\usepackage[]{hyperref}
\usepackage{hhline}
\usepackage{subcaption}
\usepackage{braket}
\usepackage{array}
\usepackage{adjustbox}
\usepackage{tabularx}

\newcolumntype{L}[1]{>{\raggedright\let\newline\\\arraybackslash\hspace{0pt}}m{#1}}
\newcolumntype{C}[1]{>{\centering\let\newline\\\arraybackslash\hspace{0pt}}m{#1}}
\newcolumntype{R}[1]{>{\raggedleft\let\newline\\\arraybackslash\hspace{0pt}}m{#1}}

\begin{document}

\begin{abstract}
The GW self-energy method has long been recognized as the gold standard for quasiparticle (QP) calculations of solids in spite of the fact that the neglect of vertex corrections and the use of a DFT starting point lacks rigorous justification. In this work we remedy this situation by including a simple vertex correction that is consistent with an LDA starting point. We analyse the effect of the self-energy by splitting it into a short-range and long-range term which are shown to govern respectively the center and size of the band gap. The vertex mainly improves the short-range correlations and therefore has a small effect on the band gap, while it shifts the band gap \emph{center} up in energy by around 0.5 eV in good agreement with experiments. Our analysis also explains how the relative importance of short- and long-range interactions in structures of different dimensionality is reflected in their QP energies. Inclusion of the vertex comes at practically no extra computational cost and even improves the basis set convergence compared to GW. The method thus provides an efficient and rigorous improvement over the GW approximation and sets a new standard for quasiparticle calculations of solids. 
\end{abstract}

\title{Simple vertex correction improves GW band energies of bulk and two-dimensional crystals}

\author{Per S. Schmidt}
\author{Christopher E. Patrick}
\altaffiliation[Present address: ]{Department of Physics, University of Warwick, Coventry CV4 7AL, United Kingdom}
\author{Kristian S. Thygesen}

\affiliation{Center for Atomic-scale Materials Design (CAMD), Technical University of Denmark, DK-2800 Kongens Lyngby, Denmark}

\date{\today}

\maketitle

\section{Introduction}
The GW approximation~\cite{aulbur_quasiparticle_1999, aryasetiawan_gw_1998,hybertsen_first-principles_1985,godby_accurate_1986}, introduced by Hedin in 1965~\cite{hedin} remains the most widely used method for quasiparticle (QP) calculations of semiconductors and insulators. Over the years it has been extensively applied to inorganic solids~\cite{shishkin_implementation_2006, kotani_all-electron_2002, marini_yambo:_2009} and more recently also to molecules~\cite{rostgaard_fully_2010, caruso_self-consistent_2013, bruneval_$gw$_2009, blase_first-principles_2011} and atomically thin two-dimensional (2D) materials~\cite{falco, filip, louie_2d}. 

The GW self-energy can be obtained by iterating Hedin’s equations once starting from $\Sigma=0$ (i.e. the Hartree approximation). This produces the trivial vertex function $\Gamma(1,2,3)=\delta(1,2)\delta(1,3)$, which corresponds to invoking the time-dependent Hartree approximation for the dynamical screening (i.e. the random phase approximation (RPA)). For this approach to be consistent, the Green's function which should be used for the calculation of the self-energy is the Hartree $G$. This is known to be a poor approximation, and instead practical GW calculations follow a “best $G$, best $W$” philosophy~\cite{hybertsen_first-principles_1985}. Most often one uses a non-interacting $G_0$ from density functional theory (DFT) and evaluates $W$ within the RPA from the polarisability $\chi_0=G_0 G_0$. This approximation is referred to as G$_0$W$_0$ and has shown to yield reasonably good, although somewhat underestimated, band gaps~\cite{shishkin_implementation_2006, Galli}. Carrying out self-consistency in the Green's function only, GW$_0$, has been found to improve the band gaps~\cite{vasp_2007}. Iterating to full self-consistency in both the Green's function and screened interaction, GW, systematically overestimates the band gaps and worsens the agreement with experiments~\cite{vasp_2007}. 

The most obvious way to go beyond the GW approximation is to perform another iteration of Hedin’s equations starting from $\Sigma = iGW$. Neglecting derivatives of $W$ this produces the kernel $\delta \Sigma(1,2)/\delta G(3,4) =iW(1,2,3,4)$, which is known from the Bethe-Salpeter Equation. The four-point nature of this kernel makes it difficult to invert the vertex equation, $\Gamma = \delta + K G G \Gamma$, without loss of accuracy. Instead one can perform a single iteration of the vertex equation to obtain $\Gamma=\delta+ WGG$, which leads to a self-energy consisting of a second-order screened exchange term in addition to the usual $iGW$ term. Gruneis \emph{et al.} have shown, using a static approximation for $W$ in the vertex, that this GW$\Gamma^1$ approximation, performed in a fully self-consistent manner, leads to significant improvements for band gaps and ionization potentials of solids~\cite{gruneis}. From a theoretical point of view this is a highly satisfactory result. The drawback is the higher complexity of the formalism and the concomitant loss of physical transparency as well as the significant computational overhead as compared to the GW method.  

Time-dependent density-functional theory (TDDFT) in principle offers a framework for including exchange-correlation (xc)-effects in the dynamical response via a two-point vertex function rather than the computationally challenging three-point vertex function that arises naturally in the diagrammatic many-body formalism. While it appears attractive to use TDDFT derived vertex functions for many-body calculations, progress along these lines has been hindered by the poor quality of the local xc-kernels derived from standard local xc-potentials. However, recent work has shown that a simple renormalization of the adiabatic LDA xc-kernel can overcome these problems and yield a dramatic improvement over the RPA for total energy calculations based on the adiabatic connection fluctuation dissipation theorem (ACDFT)~\cite{thomas, thomas2, chris}. 

Here we show that the renormalized adiabatic LDA (rALDA) kernel, when introduced in Hedin’s equations, produces a simple two-point vertex function that leads to systematically improved QP energies for a range of semiconductors and insulators. The most striking effect of the vertex is that it raises the absolute QP energies from G$_0$W$_0$ by around 0.5 eV while the gaps are almost unaffected. These effects can be traced to an improved description of the short range correlation hole and thus the (absolute) correlation energy of electrons in the ground state.
  
\section{Self-energy and xc-kernel}
As originally observed by Hybertsen and Louie~\cite{hybertsen_first-principles_1985}, it is possible to start the iterative solution of Hedin’s equation not with $\Sigma=0$ (which leads to the GW approximation), but rather with a local xc-potential: $\Sigma^0(1,2)=\delta(1,2)v_{xc}(1)$. As shown by Del Sole \emph{et al.}~\cite{delsole} this leads to a self-energy of the form 
\begin{equation}\label{eq.gw}
\Sigma(1,2) = i G(1,2) \widetilde W(1,2),
\end{equation} 
where 
\begin{equation}\label{eq.tildeW}
\widetilde W = v [1-\chi^0(v+f_{xc})]^{-1}
\end{equation} 
and $f_{xc}(1,2)=\delta v_{xc}(1)/\delta n(2)$ is the adiabatic xc-kernel. Crucially, $\widetilde W(1,2)$ is the screened \emph{effective} potential at 2 generated by a charge at 1. It consists of the bare potential plus the induced Hartree and xc-potential. It is thus the potential felt by a (Kohn-Sham) electron in the system. For comparison the potential felt by a classical test charge is the bare potential screened only by the induced Hartree potential:       
\begin{equation}
\widehat{W} = v + v[1-\chi^0(v+f_{xc})]^{-1}\chi^0 v
\end{equation} 
Using $\widehat{W}$ in Eq. \eqref{eq.gw} corresponds to including the vertex in the polarisability $P$ (or irreducible response function) but neglecting it in the self-energy. We shall refer to the use of $\widetilde W$ or $\widehat{W}$ in Eq. \eqref{eq.gw} as G$_0$W$_0\Gamma$ and G$_0$W$_0$P, respectively. The subscripts indicate that the self-energies are evaluated non-self-consistently starting from DFT. Note that in contrast to the GW approximation, which strictly should be based on the Hartree $G$, the use of a DFT starting point is perfectly justified within the G$_0$W$_0\Gamma$ theory. 

The adiabatic LDA kernel is given by 
$$ f_{xc}^\text{ALDA}[n](\mathbf{r},\mathbf{r}') = \delta(\mathbf{r}-\mathbf{r}') f_{xc}^\text{ALDA}[n]$$
where 
$$f_{xc}^\text{ALDA}[n] = \frac{d^2}{dn^2} \bigg( ne_{xc}^\text{HEG} \bigg) \bigg\vert_{n=n(\mathbf{r})},$$ 
While $f_{xc}^\text{ALDA}$ equals the exact xc-kernel of the homogeneous electron gas (HEG) in the $q\to 0$ and $\omega \to 0$ limits, it violates a number of other exact conditions. In particular, it does not incorporate the correct asymptotic $\propto q^{-2}$ behaviour for large $q$. This deficiency leads to a divergent on-top correlation hole~\cite{Furche}. Moroever, the ALDA kernel diverges at small densities where $f_{x}^\text{ALDA} \sim n^{-2/3}$. We have observed that when applying the ALDA kernel to systems other than silicon (which was the system addressed by Del Sole \emph{et al.}~\cite{delsole} and again by R. Shaltaf \emph{et al.}~\cite{shaltaf}), these divergences make it impossible to obtain converged results in practice. This is exemplified in Fig. \ref{fig:conv}\subref{subfig:conv2} and emphasizes the critical importance of renormalizing the local kernel. 

The rALDA kernel is defined for the HEG by setting $f^{\text{rALDA}}_{xc}[n](q)=f^{\text{ALDA}}_{xc}[n]$ for $q<2k_F[n]$ and $-v(q)$ otherwise (this ensures continuity at $q=2k_F$). This results in a non-local kernel with the (almost) exact asymptotic $q\to \infty$ behaviour and without the divergences of the ALDA kernel~\cite{chris}. In this work we have followed the wave vector symmetrization scheme (see Eq. (38) in  \onlinecite{chris}) to generalize the rALDA to inhomogeneous densities. Furthermore, we include only the dominant exchange part of the kernel. We mention that a small inconsistency of our QP scheme is that we iterate Hedin’s equations from $\Sigma^0(1,2)=\delta(1,2)v_{xc}^{\text{LDA}}(1)$ while $f_{xc}^{\text{rALDA}}$ does not exactly equal $\delta v_{xc}^{\text{LDA}}/\delta n$ due to the truncation at $q=2k_F$.  

The rALDA kernel provides an essentially exact description of the correlation hole of the HEG across a wide range of densities and has been shown to improve the RPA description of bond energies in molecules and solids~\cite{thomas, thomas2, chris}. However, more important for the present work is that the rALDA kernel provides a dramatic improvement of absolute correlation energies compared to RPA. For example, the RPA correlation energy of the HEG is 0.3-0.5 eV/electron too negative while the rALDA error is below 0.03 eV/electron. Very similar trends are seen for small atoms and molecules~\cite{thomas2} as well as for bulk silicon~\cite{chris}.  

\section{Computational details}
We have calculated the QP band gaps, ionization potentials (IP) and electron affinities (EA) for a range of semiconductors and insulators using five different approximations to the self-energy: (i) conventional G$_0$W$_0$ (ii) eigenvalue self-consistent GW$_0$ (iii) full eigenvalue self-consistent GW (iv) G$_0$W$_0\Gamma$ and (v) G$_0$W$_0$P. The non-self-consistent calculations employed an LDA starting point and the exchange only rALDA kernel was used to obtain $\widetilde W$ from Eq. \eqref{eq.tildeW}. 
The QP calculations for the bulk and 2D crystals in their experimental geometries were performed using the GPAW code~\cite{gpaw_enko}. See Table \ref{tab:bulkstructs} and \ref{tab:2dstructs} for lattice constants and thickness of the 2D materials.
Response functions and screened interactions were evaluated along the real frequency axis using a non-linear frequency grid. The number of unoccupied bands included in $\chi_0$ was set equal to the number of plane wave basis functions. A $8\times8\times8$ ($18 \times 18$) \textbf{k}-point grid was used for all bulk (2D) materials. For the 2D materials we employed a recently developed method for treating the critical $\mathbf q \to \mathbf 0$ limit of the screened interaction while avoiding spurious interactions with neighbouring supercells~\cite{filip}. $15\,\mathrm{\AA}$ of vacuum was used in the out-of-plane direction. The reported band positions and gaps were extrapolated to the infinite plane wave basis limit and the results are estimated to be converged to within 0.02 eV.
Band edge positions with respect to vacuum were determined by aligning the Hartree potential at the nuclei in the bulk calculations to that inside a slab with surface orientation and reconstruction as reported in available experimental studies. The considered surfaces are (111) $2\times1$ for Si in the diamond structure, (100) for MgO and LiF in the rocksalt structure and (110) for the rest of the compounds in the zinc-blende structure. The slabs are represented by 10 layers (rocksalt), 14 layers (zinc-blende) and 24 layers (diamond). The surfaces were relaxed with the PBE xc-functional~\cite{pbe}, rescaled to the experimental lattice constant, and recalculated with LDA.

\begin{table}[h]
\centering
\begin{tabularx}{\columnwidth}{L{1.5cm}L{2cm}C{3cm}r}
\hline\hline \noalign{\smallskip}
& Structure & Latt. const. (\AA) & \multicolumn{1}{c}{\textbf{k}-points} \\
\hline\noalign{\smallskip}
MgO & rocksalt & 4.212 & $8\times 8\times 8$ \\
SiC & zincblende & 4.350 & $8\times 8\times 8$ \\
LiF & rocksalt & 4.024 & $8\times 8\times 8$ \\
CdS & zincblende & 5.818 & $8\times 8\times 8$ \\
Si  & diamond & 5.431 & $8\times 8\times 8$ \\
C   & diamond & 3.567 & $8\times 8\times 8$ \\
BN  & zincblende & 3.615 & $8\times 8\times 8$ \\
AlP & zincblende & 5.451 & $8\times 8\times 8$ \\
\hline\hline
\end{tabularx}
\caption{The bulk crystal structures considered in this study. The lattice constants and \textbf{k}-point grids applied in the quasiparticle calculations are shown. }\label{tab:bulkstructs}
\end{table}

\begin{table}[h]
\centering
\begin{tabularx}{\columnwidth}{>{\hsize=.5\hsize}X>{\hsize=.3\hsize}Xccc}
\hline\hline \noalign{\smallskip}
&  & Latt. const. (\AA) & Thickness (\AA) & \textbf{k}-points \\
\hline\noalign{\smallskip}
MoS$_2$ & 2H & 3.160 & 3.170 & $18\times 18\times 1$ \\
MoSe$_2$ & 2H & 3.289  & 3.340 & $18\times 18\times 1$ \\
WS$_2$ & 2H & 3.153 & 3.360 & $18\times 18\times 1$ \\
\hline\hline
\end{tabularx}
\caption{The 2D crystal structures considered in this study. Lattice constant, layer thickness and \textbf{k}-point grids are shown. }\label{tab:2dstructs}
\end{table}

\section{Results}
\begin{table}[h]
\centering
\begin{tabularx}{\columnwidth}{lcccccc}
\hline\hline \noalign{\smallskip}
& LDA & G$_0$W$_0$ & GW$_0$ & \multicolumn{1}{c}{G$_0$W$_0$P$_0$} & \multicolumn{1}{c}{G$_0$W$_0\Gamma_0$} & Exp.\\
\hline\noalign{\smallskip}
MgO & 4.68 & 7.70 & 8.16 & 7.10 & 7.96 & 7.98 \\
CdS & 0.86 & 1.76 & 2.27 & 1.84 & 1.87 & 2.48 \\
LiF & 8.83 & 14.00 & 14.75 & 13.25 & 14.21 & 14.66 \\
SiC & 1.31 & 2.54 & 2.72 & 2.38 & 2.57 & 2.51 \\
Si & 0.52 & 1.23 & 1.34 & 1.16 & 1.29 & 1.22 \\
C & 4.10 & 5.74 & 5.97 & 5.62 & 5.69 & 5.88 \\
BN & 4.36 & 6.54 & 6.81 & 6.27 & 6.60 & 6.6 \\
AlP & 1.44 & 2.48 & 2.67 & 2.34 & 2.51 & 2.47 \\
\hline\noalign{\smallskip}
ML-MoS$_2$ & 1.71 & 2.47 & 2.61 & 2.28 & 2.47 & 2.50 \\
ML-MoSe$_2$ & 1.43 & 2.08 & 2.23 & 1.99 & 2.07 & 2.31 \\
ML-WS$_2$ & 1.33 & 2.75 & 3.07 & 2.56 & 2.81 & 2.72\\
\hline\noalign{\smallskip}
MAE & \phantom{-}1.89 & \phantom{-}0.20 & 0.17 & \phantom{-}0.41 & \phantom{-}0.16 & - \\
MSE & -1.89 & -0.19 & 0.12 & -0.41 & -0.12 & - \\
\hline \hline\noalign{\smallskip}
\end{tabularx}
\caption{\label{tab:res} Band gaps obtained using different self-energy approximations (see text). Experimental values are from \cite{shishkin} and the references therein and corrected for zero-point motion (MgO: 0.15 eV, CdS: 0.06 eV, LiF: 0.46 eV, SiC: 0.11 eV, Si: 0.05 eV, C: 0.40 eV, BN: 0.26 eV, AlP: 0.02 eV) as found in \cite{fxc_bootstrap,zpr40} and the references therein. The experimental values for the 2D materials have not been corrected for zero-point motion since only a value for MoS$_2$ was found in the literature (75 meV)\cite{mos2zpr}. Calculated values for the 2D materials include spin-orbit coupling.}
\end{table}

\begin{figure*}
\includegraphics[width=1.9\columnwidth]{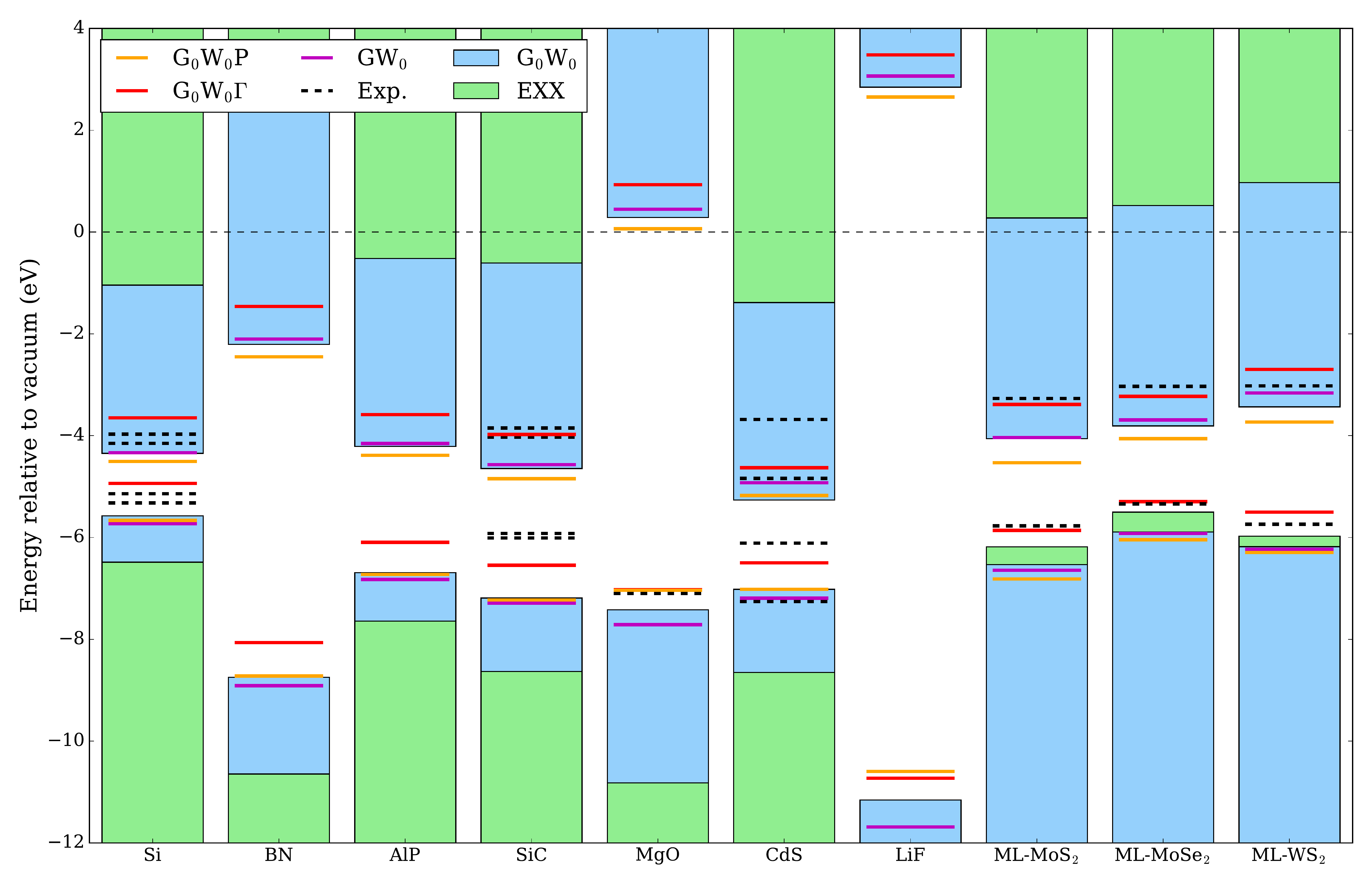}
\caption{\label{fig:IP_EA} IP and EA of a range of 3D and 2D semiconductors calculated with EXX (green), G$_0$W$_0$ (blue), G$_0$W$_0\Gamma$ (red), G$_0$W$_0$P (orange), GW$_0$ (magenta) and compared with experimental values where available (black)~\cite{IP_EA_exp}.}
\end{figure*} 

Table \ref{tab:res} shows the band gaps obtained with the different methods and their deviations from experimental reference values for both bulk and 2D materials. The 2D materials are included because they are scientifically interesting but also because they offer a unique opportunity for obtaining accurate energy levels due to their well-defined surface structures. On the contrary, energy levels in bulk solids are greatly affected by variations and uncertainties in the surface orientation/termination which makes a comparison between theoretical and experimental results problematic. 

In agreement with previous findings G$_0$W$_0$ underestimates the experimental band gaps slightly while GW$_0$ generally overestimates the gaps. The overestimation becomes even larger in GW (see Appendix) which is therefore not considered further in this work. 
The experimental VBM are from reference \cite{taisuke_acsnano}. The band gap of MoSe$_2$ is from \cite{MoSe2gapnature} where a gap of 2.18 $\pm$ 0.04 eV is reported for MoSe$_2$ on top of bilayer graphene. The effect of the substrate is calculated in the same reference to be a lowering of the band gap of 0.13 eV, giving a band gap of 2.31 eV for free-standing MoSe$_2$. \\
The band gap of 2.5 eV for free-standing MoS$_2$ is from \cite{MoS2gapnature}. In \cite{2dgapsnanoletter} a band gap of 2.18 $\pm$ 0.05 eV for MoS$_2$ on top of quartz is reported. Comparing the two numbers, quartz is expected to lower the gap by 0.32 eV. In \cite{2dgapsnanoletter} the band gap of WS$_2$ on top of quartz is reported to be 2.40 $\pm$ 0.06 eV. Assuming the same substrate effect, the band gap of free-standing WS$_2$ is 2.72 eV.\\
The numbers in Table \ref{tab:res} are including spin-orbit corrections. These are a splitting of the VB by 0.15, 0.19, 0.45 eV and of the CB by 0.00, 0.02, 0.02 eV for MoS$_2$, MoSe$_2$ and WS$_2$ respectively \cite{filip_compdata}.

From Table \ref{tab:res} we conclude that G$_0$W$_0\Gamma$ shows the best agreement with experiments, closely followed by GW$_0$, but the mean signed error of the two methods come with opposite sign. Including the vertex only in the polarizability (G$_0$W$_0$P) leads to a closing of the gap (as previously reported in \cite{fxc_bootstrap, shishkin}) resulting in significantly underestimated gaps.     

In Fig. \ref{fig:IP_EA} we show the absolute positions of the valence band maximum (VBM) and conduction band minimum (CBM) relative to vacuum. The most striking effect of including the vertex is a systematic upshift of the band edges by around 0.5 eV. Remarkably, this upshift leads to a better overall agreement with experiments (dashed black lines). The upshift of band energies is not observed when the vertex is included only in the polarisability, i.e. when employing a test charge-test charge screened interaction (G$_0$W$_0$P). Moreover, no systematic upshift of the band edges is observed for the self-consistent GW flavours which also employ test charge-test charge screening. We conclude that the upshift of band energies originates from the presence of the vertex in the self-energy, i.e. the use of a test charge-electron screened interaction. 

\section{Discussion}
In the following we analyse our results from a total energy perspective focusing on the G$_0$W$_0$ and G$_0$W$_0\Gamma$ methods by a generalization of Koopmans' theorem. Subsequently the effect of short- and long-range screening is discussed. It is then exemplified how the vertex affects the results depending on if its included in the polarizability, self-energy or both. Finally the improved numerical convergence behaviour upon inclusion of the kernel is shown.

\subsection{Generalized Koopmans' theorem}
\begin{figure*}[t!]
\centering
\includegraphics[clip, trim=0.5cm 11.cm 0.5cm 1.5cm, width=1.7\columnwidth]{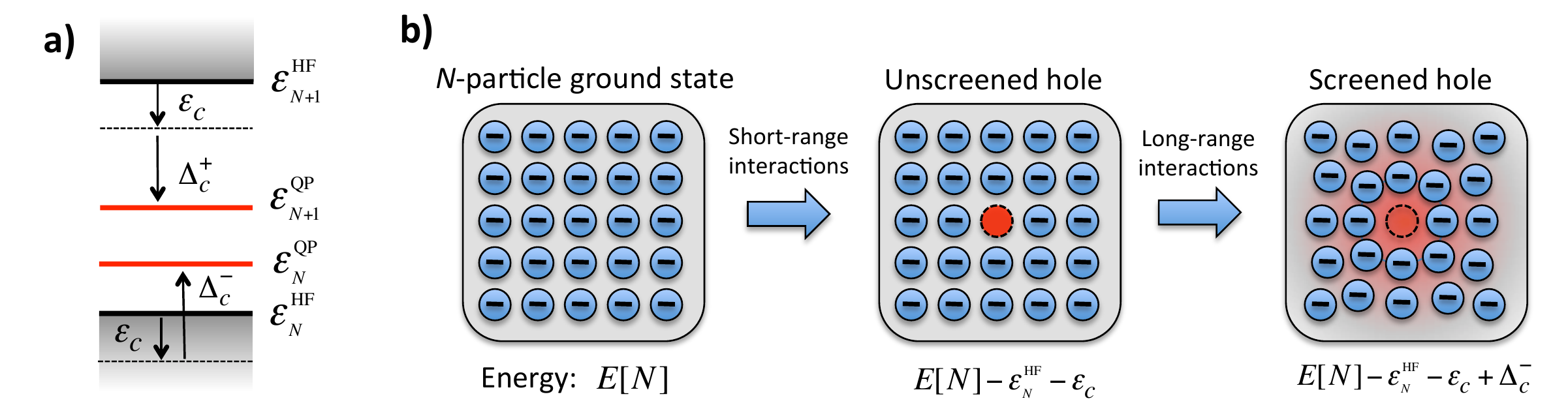}
\caption{\label{fig:diagram}(a) Schematic illustration of the different contributions to the highest occupied and lowest unoccupied QP levels of a semiconductor. (b) The energy cost of removing a valence electron consists of the Hartree-Fock energy ($\varepsilon_N^{\text{HF}}$), the correlation energy of an electron in the ground state ($\varepsilon_c$), and a stabilising screening contribution ($\Delta_c^\pm$). The two latter are predominantly of short-range and long-range nature, respectively.}
\end{figure*} 

From Koopmans’ theorem it follows that the highest occupied and lowest unoccupied QP energies can be expressed as
\begin{eqnarray}\label{eq.qp1exact}
\varepsilon^{\text{QP}}_{N} &=& \varepsilon^{\text{HF}}_{N} + E_{c}[N]-E_c[N-1] \\ \label{eq.qp2exact}
\varepsilon^{\text{QP}}_{N+1} &=& \varepsilon^{\text{HF}}_{N+1} + E_{c}[N+1]-E_c[N]
\end{eqnarray}
where $\varepsilon^{\text{HF}}$ are the Hartree-Fock single particle energies (evaluated on Kohn-Sham orbitals) and $E_c[N]$ is the correlation energy of the $N$-particle ground state. The latter can be calculated from the ACDFT, which can be cast in the form   
\begin{equation}\label{eq.acdft}
E_c = -\int_0^1 d \lambda\int_0^{\infty} \frac{d \omega}{2\pi} \text{Tr}\{ \chi^0(i\omega) (\widetilde W^\lambda(i\omega)-v)\}
\end{equation}
Here $\widetilde W^{\lambda}$ equals the screened test charge-electron interaction of Eq. (\ref{eq.tildeW}). Setting $f_{xc}=0$ we have $\widetilde W = W$ and $E_c$ becomes the RPA correlation energy.
Assuming no orbital relaxations (which is justified for an extended periodic crystal), Niquet \emph{et al.}~\cite{niquet} have shown that the ionization potential (IP) and electron affinity (EA) calculated as total energy differences with the ACDFT-RPA equal the highest occupied and lowest unoccupied QP energies from G$_0$W$_0$, respectively (when setting the renormalization factor $Z$ to unity). In the same way it can be shown, at least for an exchange only kernel, that the IP and EA obtained from the ACDFT with $\widetilde W$ from Eq. (\ref{eq.tildeW}), equal the respective QP band edges obtained from G$_0$W$_0\Gamma$ when $\Gamma$ is the vertex corresponding to $f_x$ (see the Appendix for a proof). These results represent a generalization of Koopmans’ theorem of Hartree-Fock theory.   

In general, HF is known to significantly overestimate the band gap of solids (see Fig. \ref{fig:IP_EA}). Comparing with Eqs. (\ref{eq.qp1exact}-\ref{eq.qp2exact}) this means that the correlation energy in the $N \pm 1$ states must be larger (more negative) than the correlation energy in the neutral $N$-particle ground state. It might seem surprising that $E_c[N-1]<E_c[N]$ since naively one would expect the correlation energy to be a monotonic decreasing function of $N$. However, the addition of an electron/hole to the system changes its character from insulating to metallic and this entails an increase in the correlation energy. To make this idea more explicit we can split the change in correlation energy into two terms: the correlation energy per electron in the neutral ground state ($\varepsilon_c\equiv E_c[N]/N<0$) and a remainder representing the extra correlation energy due to the insulator-metal transition ($\Delta_c^{+/-} \equiv E_c[N\pm 1]-(E_c[N]\pm \varepsilon_c)$). With these definitions we can write 
\begin{eqnarray}\label{eq.qp1}
\varepsilon^{\text{QP}}_{N} &=& \varepsilon^{\text{HF}}_{N} + \varepsilon_c-\Delta_c^- \\ \label{eq.qp2}
\varepsilon^{\text{QP}}_{N+1} &=& \varepsilon^{\text{HF}}_{N+1} + \varepsilon_c+\Delta_c^+
\end{eqnarray}   

The relations are illustrated in Fig. \ref{fig:diagram}(a). Clearly, the effect of $\varepsilon_c$ is to downshift the band edges from their HF positions while the $\Delta_c^{\pm}$ closes the gap. In the quasiparticle picture, $\Delta_c^{\pm}$ represent the screening of the added electron/hole, see Fig. \ref{fig:diagram}(b), and we shall therefore refer to them as screening terms. By its stabilization of the final states (the $N\pm 1$ states) the effect of the screening terms is similar to that of orbital relaxations in finite systems, yet the underlying physics is completely different: orbital relaxations are vanishingly small in periodic crystals and occur even in non-correlated theories like HF. In contrast $\Delta_c^{\pm}$ describes a pure correlation effect and does not vanish in infinite, periodic systems.

\begin{figure}
\includegraphics[width=1.\columnwidth]{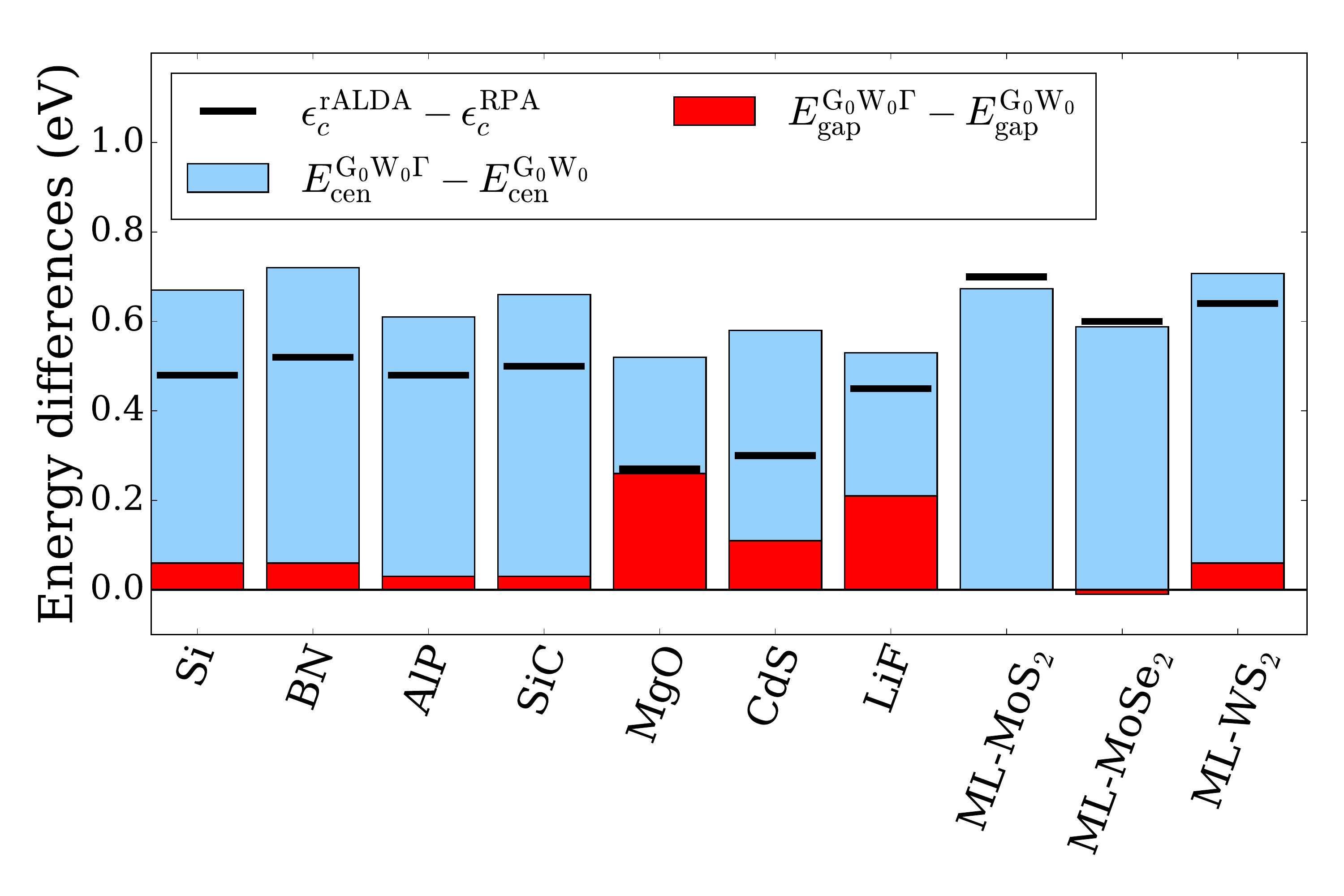}
\caption{\label{fig:epsilon_c} The difference in band gap, band gap center, and $\epsilon_c$ upon inclusion of the rALDA kernel.}
\end{figure}

We find it useful to analyse the QP energies in terms of the band gap and the band gap center. These are related to $\varepsilon_c$ and $\Delta_c^{\pm}$ by
\begin{eqnarray}\label{eq.gap1}
E^{\text{QP}}_{\text{gap}} &=& E^{\text{HF}}_{\text{gap}} + (\Delta_c^- + \Delta_c^+)\\ \label{eq.gap2}
E^{\text{QP}}_{\text{cen}} &=& E^{\text{HF}}_{\text{cen}} + \varepsilon_c+ (\Delta_c^+-\Delta_c^-)/2
\end{eqnarray}   
The correlation contribution to the gap is determined only by the screening terms $\Delta_c^{\pm}$. From the close agreement between the G$_0$W$_0$ and G$_0$W$_0\Gamma$ band gaps (red columns in Fig. \ref{fig:epsilon_c}) we conclude that the vertex has little effect on the (sum of the) screening terms. In contrast, the band gap center also depends on the ground state correlation energy, $\varepsilon_c$. We have calculated $\varepsilon_c$ for all the investigated materials using the RPA and rALDA total energy methods (see the Appendix for the exact values). In Fig. \ref{fig:epsilon_c} we compare the difference between the RPA and rALDA calculated $\varepsilon_c$ (black line) with the difference between the G$_0$W$_0$ and G$_0$W$_0\Gamma$ calculated band gap centers (blue columns). The rather close agreement shows that the main difference in band gap center can be ascribed to $\varepsilon_c$. It is now clear that the upshift of QP energies obtained with G$_0$W$_0\Gamma$ originates from the smaller (less negative) correlation energy of electrons in the neutral ground state predicted by rALDA compared to RPA. The well documented superiority of the rALDA over the RPA for the description of ground state correlation energies, in combination the improved agreement with experimental band energies (Fig. \ref{fig:IP_EA}) constitutes strong evidence that our QP-rALDA scheme represents a genuine improvement over the GW approximation.  

\subsection{Short- versus long-range screening}
We have seen that the dominant effect of the vertex correction is to shift the band gap center while the band gap itself is less affected. Physically, the main effect of the rALDA kernel is to modify the effective Coulomb interaction at short distances. More precisely, given a density variation, $\delta n$, the corresponding induced electron potential, $\delta v_{\text{Hxc}} = (v+f_{xc})\delta n$, is generally weaker than the bare Hartree potential $\delta v_H = v \delta n$, because $v(q)+f_{xc}(q)<v(q)$. However, by definition of the rALDA kernel, the reduction is stronger for larger $q$, which translates to shorter distances in real space. From these observations we can conclude that the QP band gap is mainly determined by long-range interactions while the band gap center is sensitive to the short-range interactions. This agrees well with the quasiparticle picture illustrated in Fig. \ref{fig:diagram}: Namely, adding a particle/hole without accounting for the screening represents a local (short-range) perturbation while the screening of the added charge is a long-range process.  While the rALDA kernel mainly reduces the short-range interactions it also reduces the long-range components slightly. This leads to a slightly weaker long-range screening (smaller $\Delta_c$) and slightly larger band gaps as seen in Table \ref{tab:res}. 

Returning to Fig. \ref{fig:IP_EA} we note that for the 2D materials Hartree-Fock predicts a lower IP than the GW methods in clear contrast to the situation for bulk solids. This anomalous behaviour is a result of the relatively more important effect of short- compared to long-range correlations in reduced dimensions. Indeed, the dielectric function of a 2D semiconductor approaches unity in the long wavelength limit, which reduces the screening terms $\Delta_c^{\pm}$. At the same time we find that the 2D materials present the largest values for $\varepsilon_c$ of all the materials (see Table V in the Appendix) showing that the absolute correlation energy per electron is larger for the 2D materials compared to the bulk materials. 

\subsection{Vertex in the polarizability and/or self-energy}
\begin{figure}
\includegraphics[width=1.\columnwidth]{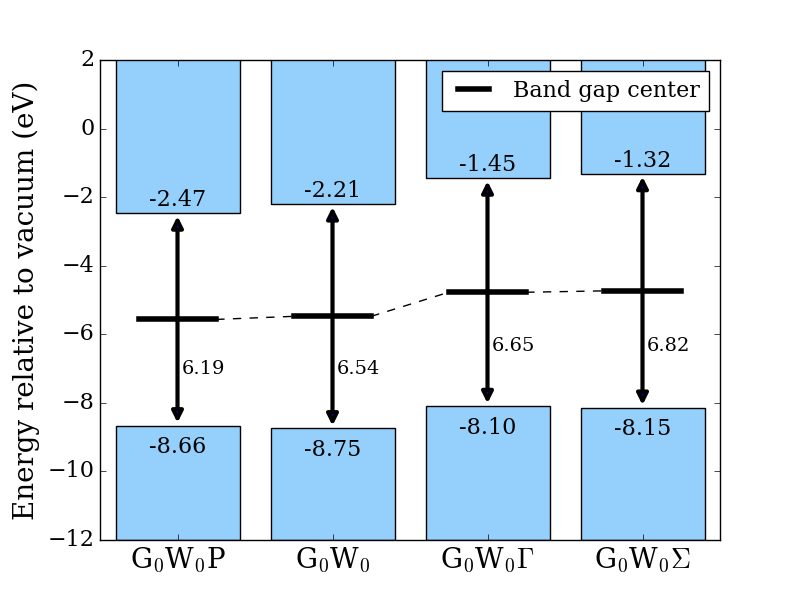}
\caption{\label{fig:BN_4methods}Absolute position of the VBM and CBM relative to vacuum for BN calculated with the four different methods. The band gap center is shown with a dotted line.}
\end{figure}
As mentioned previously, the vertex enters Hedin’s equations at two places, namely the polarisability and the self-energy. For a consistent description the vertex should be included in both places (the G$_0$W$_0\Gamma_0$ approach). However, it is instructive to study the effect of including the vertex separately in the polarisability and self-energy. To this end we note that self-energy (excluding the exchange part) can be written 
\begin{equation}
\Sigma = iG v \chi (v+f_{xc}) 
\end{equation}
where $\chi = P+Pv\chi$ and $P=\chi^0+\chi^0 f_{xc} P$. Including $f_{xc}$ in $P$ affects the description of screening while including it in $\Sigma_c$ affects the form of the potential created by the induced density when subject to a test-charge (as explained above it mainly reduces the short range part of the potential). Based on this we can obtain four different GW-like self-energies, where the only one not mentioned up to now is including the vertex in the self-energy but not in the polarisability - this we term GW$\Sigma$. 
Fig. \ref{fig:BN_4methods} shows the band gap size and center obtained for BN with these four self-energies. It is clear that the band gap center depends mainly on the $f_{xc}$ in the self-energy, i.e. a correct description of the band gap center requires the inclusion of xc-effects in the induced potential. The size of the band gap increases in the order G$_0$W$_0$P, G$_0$W$_0$, G$_0$W$_0\Gamma$, G$_0$W$_0\Sigma$. 
As previously argued the size of the gap depends mainly on the long range screening (and less on the short-range form of the final induced potential). The $f_{xc}$ reduces the long range interactions somewhat. Thus the total induced potential will generally be smaller when xc-effects are included in the final potential. This explains the larger band gaps found for G$_0$W$_0\Gamma$ and G$_0$W$_0\Sigma$. The remaining ordering comes from noting that $\chi^{\text{rALDA}}>\chi^{\text{RPA}}$ because the higher order diagrams, which reduce the effect of $\chi^0$, are larger in RPA.   

\begin{figure*}[ht!]
\centering
\begin{subfigure}[b]{1.0\columnwidth}
\includegraphics[clip, trim=0.cm 0.cm 1.5cm 0.cm, width=\columnwidth]{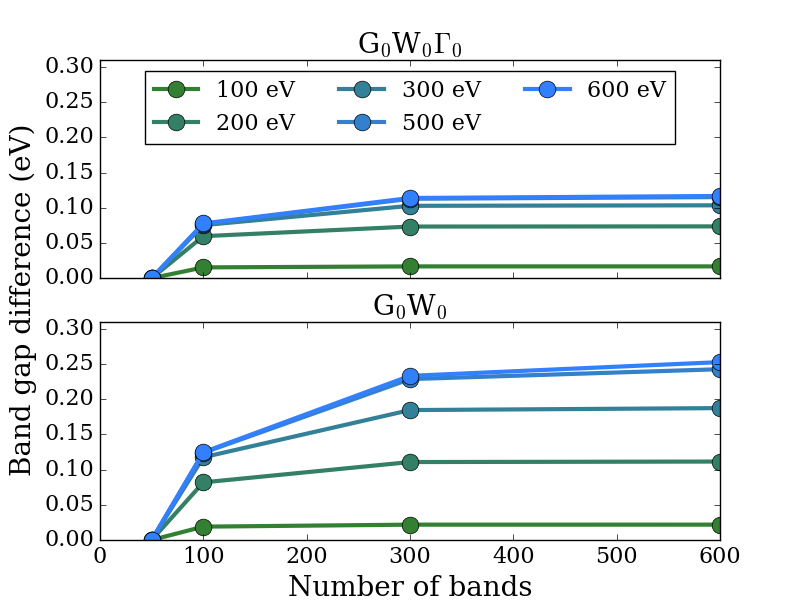}
\caption{\label{subfig:conv1}}
\end{subfigure}
\begin{subfigure}[b]{1.0\columnwidth}
  \includegraphics[clip, trim=0.cm .5cm 0.cm 0.cm,width=1.04\columnwidth]{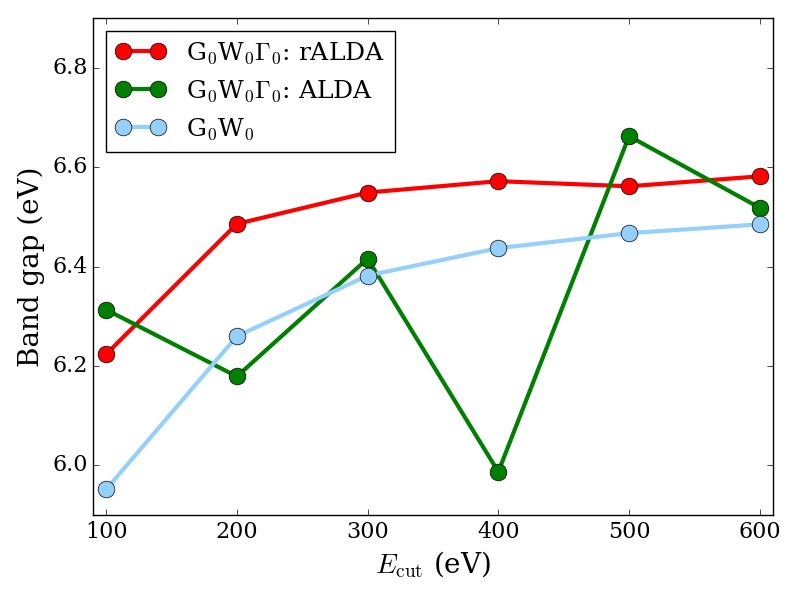}
  \caption{\label{subfig:conv2}}
\end{subfigure}
\caption{\label{fig:conv} (a) Convergence of the band gap in BN with respect to plane wave cutoff and the number of bands included using the RPA (bottom) and rALDA (top) kernel. (b) Plane wave convergence of the band gap of bulk BN using the ALDA and rALDA kernel in the G$_0$W$_0\Gamma_0$ method as well as the RPA kernel (G$_0$W$_0$ method).}
\end{figure*}

\subsection{Improved convergence} 
Finally, we mention that the reduction of large $q$ components by the rALDA kernel not only improves the description of short range correlations but also leads to faster convergence with respect to plane waves and number of unoccupied states compared to RPA/GW calculations, as shown in Fig. \ref{fig:conv}\subref{subfig:conv1} for the case of bulk BN. On the y-axis we show the difference in the band gap compared to that from a calculation at a cutoff of 50 eV with the corresponding number of bands. The improved convergence also manifests itself in the QP corrections to the individual bands. 

In Fig. \ref{fig:conv}\subref{subfig:conv2} the band gap of bulk BN is shown versus the plane wave cutoff applying the G$_0$W$_0\Gamma_0$ method with the ALDA and rALDA kernel as well as the standard G$_0$W$_0$ method. The need of the renormalization of the kernel in order to obtain converged results, is very apparent. 

\section{Conclusion}
In conclusion, we have demonstrated that a more accurate description of short-range correlations in QP calculations can be obtained with a simple TDDFT-inspired vertex function. Inclusion of the vertex improves the agreement with experimental data for the absolute band energies of bulk and two-dimensional semiconductors. Moreover, it justifies the use of DFT as a starting point for non-self-consistent QP calculations and is thus formally more rigorous than the G$_0$W$_0$@DFT approach. Importantly, these advantages come without increasing the numerical complexity or computational cost compared to G$_0$W$_0$ calculations.  

\section{Acknowledgements}
The Center for Nanostructured Graphene (CNG) is sponsored by the Danish Research Foundation, Project DNRF103.

\section{Appendix}
\appendix
\section{From rALDA total energies to GW$\Gamma$}
Using the adiabatic connection and fluctuation-dissipation theorem (ACFDT), the exact correlation energy of the system can be written in terms of the interacting response function ($\chi^\lambda(i\omega)$) of a system with a scaled Coulomb interaction, $v \rightarrow \lambda v$:
\begin{equation*}
E_c = -\int_0^1 d\lambda \int_0^\infty \frac{d\omega}{2\pi} \text{Tr} \bigg( v\bigg[ \chi^\lambda(i\omega) - \chi^\text{KS}(i\omega)\bigg] \bigg),
\end{equation*}
where $\chi^\text{KS}(i\omega)$ is the response function of the non-interacting Kohn-Sham system. $\chi^\lambda$ can in principle be obtained from the Dyson equation
$$ \chi^\lambda(i\omega) = \chi^\text{KS}(i\omega) + \chi^\text{KS}(i\omega)\bigg[ \lambda v + f_{xc}^\lambda(i\omega)\bigg] \chi^\lambda(i\omega),$$
where all the complicated correlation effects has been transferred into $f_{xc}^\lambda(i\omega)$, which needs to be approximated.\\
It can be shown, that any pure exchange kernel must have the property $f_x^\lambda[n](\mathbf{r},\mathbf{r}',i\omega) = \lambda f_x[n](\mathbf{r},\mathbf{r}',i\omega)$ making it possible to carry out the $\lambda$-integration analytically:
\begin{align*}
E_c = \int_0^\infty \frac{d\omega}{2\pi} \text{Tr} &\bigg[vf_{Hx}^{-1}(i\omega) \ln \bigg[1- \chi^\text{KS}(i\omega)f_{Hx}(i\omega)\bigg] \\
&+ v \chi^\text{KS}(i\omega)\bigg],
\end{align*}
where $f_{Hx}(i\omega) = v + f_x(i\omega)$. \\
Denoting the change in $\chi^\text{KS}(i\omega)$ when adding one electron to the lowest unoccupied KS orbital by $\delta\chi^\text{KS}(i\omega)$, 
$$ \delta \chi^\text{KS}(\mathbf{r},\mathbf{r}',i\omega) = \phi_c^*(\mathbf{r}) G_0(\mathbf{r},\mathbf{r}',\epsilon_c + i\omega) \phi_c(\mathbf{r}') + \text{c.c.},$$
assuming that the density does not change by the addition of one electron (meaning $f_{Hx}(i\omega)$ does not change), we calculate the correlation contribution to the electron affinity (AE):
\begin{align*}
AE_c &= E_c[N] - E_c[N+1] \\
&= \int_0^\infty \frac{d\omega}{2\pi} \text{Tr} \bigg[vf_{Hx}^{-1}(i\omega) \ln \bigg[1- \chi^\text{KS}(i\omega)f_{Hx}(i\omega)\bigg] \\
&+ v \chi^\text{KS}(i\omega) - \bigg[ vf_{Hx}^{-1}(i\omega) \ln \bigg[1- \bigg(\chi^\text{KS}(i\omega)\\
&+\delta \chi^\text{KS}(i\omega)\bigg)f_{Hx}(i\omega)\bigg] + v \bigg(\chi^\text{KS}(i\omega) + \delta \chi^\text{KS}(i\omega)\bigg) \bigg] \bigg] \\
&= - \int_0^\infty \frac{d\omega}{2\pi} \text{Tr} \bigg[vf_{Hx}^{-1}(i\omega) \bigg( \ln \bigg[ 1 - \chi^\text{KS}(i\omega)f_{Hx}(i\omega) \\
&- \delta\chi^\text{KS}(i\omega)f_{Hx}(i\omega)\bigg] - \ln\bigg[ 1 - \chi^\text{KS}(i\omega)f_{Hx}(i\omega)\bigg]\bigg) \\
&+ v\delta\chi^\text{KS}(i\omega) \bigg] \\
&= - \int_0^\infty \frac{d\omega}{2\pi} \text{Tr} \bigg[vf_{Hx}^{-1}(i\omega) \ln \bigg[ 1 - \delta\chi^\text{KS}(i\omega)f_{Hx}(i\omega) \\
&\bigg(1-\chi^\text{KS}(i\omega)f_{Hx}(i\omega)\bigg)^{-1} \bigg] + v\delta\chi^\text{KS}(i\omega) \bigg]
\end{align*}
By adding and subtracting $v\bigg[1-\chi^\text{KS}(i\omega)f_{Hx}(i\omega)\bigg]^{-1}\delta\chi^\text{KS}(i\omega)$, we split $AE_c$ in two terms, $AE_c = AE_c^\text{QP} + AE_c'$ and following the arguments of Niquet et al.\cite{niquet} $AE_c'$ vanishes, and left is
\begin{align*}
AE_c &= \int_0^\infty \frac{d\omega}{2\pi} \text{Tr} \bigg[ \bigg( v\bigg[1-\chi^\text{KS}(i\omega)f_{Hx}(i\omega)\bigg]^{-1} - v\bigg)\\
&\times\delta\chi^\text{KS}(i\omega) \bigg] \\
&= \int_0^\infty \frac{d\omega}{2\pi} \text{Tr} \bigg[ \bigg( W(i\omega) - v\bigg)\delta\chi^\text{KS}(i\omega) \bigg]. \\
\end{align*}
This is exactly the correlation contribution to the conduction band matrix element of the GW self-energy in the GW$\Gamma$ method:
$$ AE_c = \braket{\phi_c | \Sigma_c(\epsilon_c)|\phi_c}$$
where 
$$ \Sigma_c(\mathbf{r},\mathbf{r}',\epsilon_c) = -\int \frac{d\omega}{2\pi} G_0(\mathbf{r},\mathbf{r}',\epsilon_c + i\omega) W(\mathbf{r},\mathbf{r}',i\omega)$$
and 
$W(i\omega) = v\bigg[1-\chi^\text{KS}(i\omega)f_{Hx}(i\omega)\bigg]^{-1}$.

\section{Eigenvalue self-consistency}
The effect of eigenvalue self-consistency in G (GW$_0$) and in both G and W (GW) is shown in Table \ref{tab:resgw0}. The effect of the kernel is seen to be independent of eigenvalue self-consistency.
\begin{table}[h!]
\centering
\begin{tabularx}{\columnwidth}{lccc|cccc}
\hline\hline \noalign{\smallskip}
& GW$_0$ & \multicolumn{1}{c}{GW$_0$P$_0$} & \multicolumn{1}{c}{GW$_0\Gamma_0$} & GW & \multicolumn{1}{c}{GWP$_0$} & \multicolumn{1}{c}{GW$\Gamma_0$} & Exp.\\
\hline\noalign{\smallskip}
MgO & 8.16 & 7.52 & 8.21 & 9.21 & 8.56 & 9.47 & 7.98 \\
CdS & 2.27 & 2.03 & 2.34 & 2.73 & 2.44 & 2.83 & 2.48 \\
LiF & 14.75 & 14.02 & 14.90 & 16.29 & 15.59 & 16.46 & 14.66 \\
SiC & 2.72 & 2.52 & 2.73 & 3.06 & 2.80 & 3.11 & 2.51 \\
Si & 1.34 & 1.23 & 1.40 & 1.48 & 1.38 & 1.60 & 1.22 \\
C & 5.97 & 5.74 & 5.88 & 6.38 & 6.16 & 6.33 & 5.88 \\
BN & 6.81 & 6.50 & 6.84 & 7.44 & 7.08 & 7.49 & 6.6 \\
AlP & 2.67 & 2.48 & 2.67 & 3.00 & 2.74 & 3.01 & 2.47 \\
\hline \noalign{\smallskip}
MAE & 0.16 & 0.23 & 0.18 & 0.72 & 0.38 & 0.81 & - \\
MSE & 0.11 & -0.22 & 0.15 & 0.72 & 0.37 & 0.81 & - \\
\hline\hline
\end{tabularx}
\caption{\label{tab:resgw0} Bandgaps calculated with eigenvalue self-consistency in G (GW$_0$) and in both G and W (GW).}
\end{table}

\section{Values of $\epsilon_c$ and $\Delta_c^{\pm}$}
The absolute values of $\epsilon_c$ and $\Delta_c^{\pm}$ as well as the sum and differences of $\Delta_c^{\pm}$ contributing to the band gaps and centers respectively are printed in Table \ref{tab:resepsilon}. The RPA and rALDA total energy calculations were done with the implementation described in \cite{chris}. $8\times8\times8$ k-points were used and the correlation energy was extrapolated to an infinite plane wave cutoff from calculations up to 500 eV. The frequency integration was done using 16 frequency points on the imaginary axis together with a Gauss-Legendre quadrature. 
\begin{table*}[h]
\makebox[\textwidth][c]{
\begin{tabular}{lccc|cc|cc|ccc|ccc}
\hline\hline \noalign{\smallskip}
& \multicolumn{3}{c}{$\epsilon_c$ from total energies} & \multicolumn{2}{c}{$\Delta_c^+$} & \multicolumn{2}{c}{$\Delta_c^-$} & \multicolumn{3}{c}{$\Delta_c^- + \Delta_c^+$} & \multicolumn{3}{c}{($\Delta_c^+ - \Delta_c^-$)/2}\\
\hhline{~-------------}
\noalign{\smallskip}
& RPA & rALDA & $\Delta$ & G$_0$W$_0$ & G$_0$W$_0\Gamma_0$ & G$_0$W$_0$ & G$_0$W$_0\Gamma_0$ & G$_0$W$_0$ & G$_0$W$_0\Gamma_0$ & $\Delta$ & G$_0$W$_0$ & G$_0$W$_0\Gamma_0$ & $\Delta$\\
\hline
Si & -1.53 &  -1.05  & 0.48 & -1.77 & -1.56  & -2.44 & -2.60 & -4.21 & -4.16 & 0.05 & 0.34 & 0.52 & 0.19\\
BN & -1.75 &  -1.23  & 0.52 & -3.02 &   -2.79 & -3.65  & -3.81 & -6.67 & -6.60 & 0.07 & 0.32 & 0.51 & 0.20\\
AlP & -1.50 &  -1.02  & 0.48 & -2.19 & -2.05  & -2.45 & -2.57 & -4.64 & -4.62 & 0.02 & 0.13 & 0.26 & 0.13\\
SiC & -1.64 &  -1.14 & 0.50 & -2.40 &   -2.23 & -3.08 &  -3.23 & -5.48 & -5.46 & 0.02 &  0.34 & 0.50 & 0.16 \\
MgO & -1.00 &  -0.73  & 0.27& -2.84 &   -2.46 & -4.40 &  -4.53 & -7.24 & -6.99 & 0.25 & 0.78 & 1.04 & 0.26\\
CdS &  -1.13 & -0.83  &  0.30 & -2.75  & -2.41 & -2.76  & -2.98 & -5.51 & -5.39 & 0.12 & 0.01 & 0.29 & 0.28\\
LiF & -1.59 &  -1.14 & 0.45 & -1.87 &  -1.69 & -5.96 &   -5.94 & -7.83 & -7.63 & 0.2 & 2.05 & 2.13 & 0.08\\
\hline 
ML-MoS$_2$ & -1.97 & -1.27 & 0.70 & -2.37 & -2.39 & -1.62 & -1.60 & -3.99 & -3.99 & 0.00 & -0.38 & -0.40 & 0.02 \\
ML-MoSe$_2$ & -1.87 & -1.27 & 0.60 & -2.36 & -2.48 & -1.58 & -1.47 & -3.94 & -3.95 & 0.01 & -0.39 &  -0.51 & 0.12\\
ML-WS$_2$ & -1.73 & -1.09 & 0.64 & -2.44 & -2.40 & -1.76 & -1.74 & -4.20 & -4.14 & 0.06 & -0.34  & -0.33 & 0.01\\
\hline\hline
\end{tabular}
}
\caption{}\label{tab:resepsilon}
\end{table*}

\section{2D materials: IP, EA and gaps}
VBM, CBM and band gaps for three 2D semiconductors calculated with various methods, relative to vacuum, are shown in Table \ref{tab:2d}. The HSE, G$_0$W$_0$ and G$_0$W$_0\Gamma$ are all non-self-consistent calculations on top of LDA orbitals and eigenvalues. The procedure of obtaining the experimental values is described in the article.

\begin{table*}[h]
\centering
\begin{tabular}{lC{1.0cm}C{1.1cm}C{1.1cm}C{1.5cm}C{1.5cm}C{1.5cm}C{1.0cm}}
\hline\hline \noalign{\smallskip}
& LDA & HSE & G$_0$W$_0$ & G$_0$W$_0$P & G$_0$W$_0\Gamma$ & GW$_0$ & Exp.\\
\hline\noalign{\smallskip}
\multicolumn{8}{c}{MoS$_2$} \\
\hline\noalign{\smallskip}
VBM & -6.07 & -6.19 & -6.53 & -6.81 & -5.86 & -6.65 & -5.77\\
CBM & -4.35 & -3.93 & -4.06 & -4.53 & -3.39 & -4.04 &-3.27\\
Gap & \phantom{-}1.71 & \phantom{-}2.25 & \phantom{-}2.47 & \phantom{-}2.28 & \phantom{-}2.47 & \phantom{-}2.61 & \phantom{-}2.50\\
\hline\hline\noalign{\smallskip}
\multicolumn{8}{c}{MoSe$_2$} \\
\hline\noalign{\smallskip}
VBM & -5.48 & -5.58 & -5.89 & -6.04 & -5.30 & -5.92 & -5.34 \\
CBM & -4.06 & -3.64 & -3.81 & -4.06 & -3.23 & -3.69 & -3.03 \\
Gap & \phantom{-}1.43 & \phantom{-}1.94 & \phantom{-}2.08 & \phantom{-}1.99 & \phantom{-}2.07 & \phantom{-}2.23 & \phantom{-}2.31\\
\hline\hline\noalign{\smallskip}
\multicolumn{8}{c}{WS$_2$}\\
\hline\noalign{\smallskip}
VBM & -5.59 & -5.78 & -6.18 & -6.29 & -5.50 & -6.23 & -5.74 \\
CBM & -4.26 & -3.95 & -3.43 & -3.73 & -2.70 & -3.16 & -3.02 \\
Gap & \phantom{-}1.33& \phantom{-}1.83 & \phantom{-}2.75 & \phantom{-}2.56 & \phantom{-}2.81 & \phantom{-}3.07 & \phantom{-}2.72\\
\hline\hline
\end{tabular}
\caption{\label{tab:2d}}
\end{table*}

\bibliography{references}

\end{document}